\documentclass[11pt]{article}
\usepackage{amsmath,amsfonts}
\usepackage{times}
\usepackage[left=2.54cm,top=2.54cm,right=2.54cm,bottom=2.54cm,bindingoffset=0.0cm]{geometry}
\usepackage{setspace}
\usepackage{enumerate}
\usepackage{enumitem}
\usepackage{wrapfig}
\usepackage{fullpage}
\usepackage{color}
\usepackage{titlesec}
\usepackage[numbers,square,compress]{natbib}
\titleformat{\section} 
  {\normalfont\fontsize{12}{17}\sffamily\bfseries}
  {\thesection}
  {1em}
  {}
  
\titleformat{\subsection} 
  {\normalfont\fontsize{11}{17}\sffamily\bfseries\slshape}
  {\thesubsection}
  {1em}
  {}

\titleformat{\subsubsection}
{\color{black}\normalfont\fontsize{11}{1}\slshape}
{\color{black}\thesection}{-5em}{}

\usepackage{parskip} % skip paragraph indentations
\usepackage[left]{lineno}
\usepackage{lipsum}
\usepackage{hyperref}
\usepackage{graphicx}
\usepackage{amssymb}
\usepackage[x11names]{xcolor}
\usepackage{hyperref}
\usepackage{enumitem}
\usepackage{booktabs}
\usepackage{amsmath}
\usepackage{amssymb}
\usepackage{mathrsfs}
\usepackage[labelfont=bf]{caption} % to bold caption "Figure XX"
\usepackage{caption} 
\usepackage{framed} % to add frames around comments

\usepackage{authblk} % for title page

\usepackage{datetime2}
\usepackage[at]{easylist}% easy lists with @ starting each item

\usepackage{booktabs}

\begin{document}
%\title{What are we missing about animal social complexity?}
%\title{The importance of complex systems concepts in animal social complexity measures}
\title{Rethinking animal social complexity measures with the help of complex systems concepts}

\author[1,*]{Elizabeth A. Hobson}
\author[1,2]{Vanessa Ferdinand}
\author[1]{Artemy Kolchinsky}
\author[1]{Joshua Garland}
\affil[1]{Santa Fe Institute, 1399 Hyde Park Road, Santa Fe, NM 87501 USA}
\affil[2]{University of Melbourne, Parkville VIC 3010, Australia}
\affil[*]{Corresponding author}

\date{} 
\maketitle

\centerline{Version Mar 15 2019, re-submitted to \textit{Animal Behaviour}}
%\linenumbers
%\doublespace

%%%%%%%%%%%%%%%%%%%%%%%%%%%%%%%%%%%%%
\section*{Abstract}
%%%%%%%%%%%%%%%%%%%%%%%%%%%%%%%%%%%%%

Explaining how and why some species evolved to have more complex social structures than others has been a long-term goal for many researchers in animal behavior because it would provide important insight into the links between evolution and ecology, sociality, and cognition. However, despite long-standing interest, the evolution of social complexity is still poorly understood. This may be due in part to researchers focusing on the feasibility of quantifying aspects of sociality, rather than what features are characteristic of animal social complexity in the first place. Any given approach to studying complexity can tell us some things about animal sociality, but may miss others, so it is critical to decide first how to conceptualize  complexity before jumping in to quantifying it. Here, we briefly summarize five existing approaches to measuring social complexity. Then, we highlight three fundamental concepts that are commonly used in the field of complex systems: (1) scales of organization, (2) compression, and (3) emergence. All of these concepts are applicable to the study of animal social systems, but are not often explicitly addressed in existing social complexity measures. We discuss how these concepts can provide a rigorous foundation for conceptualizing social complexity, the potential benefits of incorporating them, and how existing measures do (or do not) include them. Ultimately, researchers need to critically evaluate any measure of animal social complexity in order to balance the biological relevance of the aspect of sociality they are quantifying with the feasibility of obtaining enough data. 

%%%%%%%%%%%%%%%%%%%%%%%%%%%%%%%%%%%%%
\textit{\textbf{Keywords:} Animal sociality, complex systems, coarse-graining, compression, downward causation, emergence, feedback, self-organization, social scale, social structure}
%%%%%%%%%%%%%%%%%%%%%%%%%%%%%%%%%%%%%

\newpage

\section*{Highlights} 

\begin{itemize}
  \item  The richness of animal social complexity is inherently challenging to quantify and current approaches fall short.
  \item Fundamental concepts from complex systems theory (scales of organization, compression, and emergence), when coupled with traditional approaches, will lead to new insights and ways of thinking about animal social complexity.
    \item Complex systems concepts may also be useful in moving beyond purely descriptive accounts of sociality and allowing practitioners to create causal accounts of animal social complexity.
\end{itemize}

%%%%%%%%%%%%%%%%%%%%%%%%%%%%%%%%%%%%%
\section*{Introduction}
%%%%%%%%%%%%%%%%%%%%%%%%%%%%%%%%%%%%%

Social systems reflect the rich landscape and diversity of social solutions that animals have evolved in order to thrive in a wide range of habitats and conditions. These solutions vary in complexity, from apparently simple ones to those that appear more complex. Explaining how and why some species evolved to have more complex social structures than others has been a long-term goal for many researchers in animal behavior because it would provide important insight into the links between evolution and ecology, sociality, and cognition. However, despite long-standing interest, the evolution of social complexity is still poorly understood~\citep{de2009animal}, largely due to disagreement about how to best quantify complexity, how to compare complexity across species, whether these broad comparisons are biologically meaningful, and what complexity is in the first place. In this paper, we critically consider which aspects of complexity we should incorporate in order to evaluate different approaches to social complexity. 

There are two major reasons to quantify animal social complexity, and these reasons can shape how social complexity is quantified. First, researchers are often drawn to making comparisons across groups, which we call the ``different is interesting'' motivation. When species systematically vary on some dimensions of their social organization, then that itself is something interesting to study, understand, and explain, and can indicate promising areas for future research. For example, if two bird species are very similar but one has group sizes of 10 while the other forms groups of over 1000, then it is intrinsically interesting to explain why those differences occur. Second, and especially relevant to social and cognitive research, is the ``more is harder'' motivation. For social systems, there is an implicit (or sometimes explicit) assumption that  higher social complexity (however it is measured) is inherently difficult to originate and maintain. This motivation is often framed in evolutionary terms, in that it is conjectured that natural selection produced special mechanisms that allows some species to form, maintain, and process more complex social systems. In this case, it is then interesting to identify those mechanisms and understand how they evolved. 

These two motivations have led to what we view as two different approaches to quantifying animal social complexity. We term these \textit{descriptive} and \textit{causal} approaches. This distinction parallels recent work that suggests that social complexity can be considered from an ``insider's'' perspective versus an ``outsider's'' perspective~\citep{Aureli2019SocialGroups}, but differs in that our distinction between descriptive and causal approaches is more general and not linked to analyses on a specific scale. Researchers driven by the ``different is interesting'' motivation generally approach measuring social systems from an ``outsider's'' perspective and quantify social complexity with a \textit{descriptive} approach, which can allow researchers to compare these systems to one another. Researchers may want to use this descriptive approach to get a useful statistic to characterize societies, or as a quantitative measure of the amount of organization or amount of pattern in the collective behavior of groups of animals to facilitate comparisons of sociality across different species. 

Alternatively, researchers driven by the ``more is harder'' motivation take more of the perspective of animals within the society, and adopt a \textit{causal} approach to animal social complexity. This causal approach is concerned with uncovering the processes or mechanisms that underlie the organization and complexity of societies. Causes of particular social patterns can range widely, across cognitive, perceptual, ecological, spatial, and physiological mechanisms. Researchers may  use a causal approach to understand what types of information play a role in structuring social organization, or to gain insight into the cognitive abilities needed to support and function within a certain kind of social system. An important test of a causal approach is to see if it leads to quantitative generative models of animal societies that can replicate and predict aspects of real empirical data. 

There is an intimate link between these two approaches to animal social complexity: a descriptive approach may allow researchers to rank systems in terms of their relative complexity, while a causal approach can identify the mechanics that cause one system to be more or less complex than another. In many cases, scientists are interested in understanding the mechanisms underlying social behavior, but cannot observe those mechanisms directly, so may instead employ descriptive measures as proxies for reasoning about the causal mechanisms. Both descriptive and causal approaches to complexity, and their associated measures, operate on a continuum, with most actual measures of complexity being motivated by considerations that fall somewhere between these two extremes. Different types of social complexity measures are required in order to address particular study goals. 

One way to explicitly evaluate and compare different aspects of animal social complexity measures, and whether they are suited for a particular study question, is to incorporate some of the fundamental concepts of complex systems theory. In the past 30 to 40 years, the field of complex systems has developed theories of how underlying microscopic interactions in distributed, multi-component systems give rise to novel properties at the macro-scale (for an approachable introduction to complexity and sociality, see~\citep{Miller2009ComplexLife}). Many of the fundamental concepts in the field of complex systems can guide thinking about the ``amount'' and ``types'' of social organization, and in this way help formalize intuitive notions of social complexity. However, most of these concepts have been largely isolated from research on animal behavior and animal sociality. A better understanding of some of the fundamental concepts in complex systems can help  critically evaluate existing approaches to animal social complexity as well as suggest promising avenues for developing new approaches.

Here, we summarize several existing measures of animal social complexity. We then highlight three foundational concepts that are commonly used in the field of complex systems: (1) scales of organization, (2) compression, and (3) emergence. We show how they can provide a rigorous foundation for approaching social complexity and how they can help circumvent some of the pitfalls of certain existing animal social complexity measures. Using these concepts from complex systems, we re-examine existing animal social complexity measures and show how they relate to these fundamental concepts.

\section*{Overview of social complexity measures}
Complex social systems are comprised of individuals that interact with many other individuals across different social contexts and over time (adapted from~\citep{Freeberg2012SocialComplexity.}, see also~\citep{Kappeler2019AComplexity}). Methods for studying animal social complexity vary widely in their approaches and are often divided into whether they focus on social relationships (more common for vertebrate systems) or on social organization and social roles (more common for social insect systems)~\citep{Weiss2019MeasuringModels, Lukas2018}. The development of  animal social complexity measures has been driven largely by what can be feasibly measured in animal societies. While any given approach to complexity can tell us some things about animal sociality, it is critical to recognize that any particular measure may miss other aspects that may also be important.

Behavioral ecologists have quantified five main aspects of animal social complexity (partially summarized in~\citep{Kappeler2019SocialEvolution}): (1) \textit{group size}, the number of individuals in the society, (2) \textit{social roles}, the number, types, and diversity of roles individuals take on in groups, (3) \textit{levels of structure}, the extent to which groups are organized across different social scales, (4) \textit{relationship differentiation}, the extent to which relationships are individualized and differentiated, and (5) \textit{social uncertainty}, the degree to which social situations and interaction outcomes are predictable or uncertain. In this section, we briefly summarize these approaches, so as to provide context for considering how concepts from complex systems are (or are not) incorporated into each animal social complexity measure.

\subsubsection*{Group size} 
Animal social complexity can be measured in terms of group size (the average number of individuals present in a social grouping,~\citep{Dunbar1995NeocortexHypothesis,dunbar1998social}) or network size (the number of individuals known to each other,~\citep{Kudo2001NeocortexPrimates}). These measures assume that social complexity is dependent on the number of others that an individual must recognize and remember: the more individuals to recognize and remember, the more cognitively taxing the social system is assumed to be. Thus, although group size is a descriptive measure, it has a causal motivation. The availability of group size data across a wide range of species enabled early comparative analyses~\citep{dunbar1998social,Dunbar1995NeocortexHypothesis}, but mainly focused on vertebrate social complexity rather than other groups like social insects~\citep{Kappeler2019AComplexity}. The biological relevance of interspecific comparisons, and in particular the use of group size as a measure of complexity, has been strongly debated~\citep{Dunbar2009TheEvolution,Shultz2010SocialInvestment,Shultz2007TheVertebrates.}.

\subsubsection*{Levels of structure} 
Social complexity can be quantified as the extent to which social groups can be organized into nested layers or hierarchical levels~\citep{Whitehead2008,Wittemyer2005TheStructures}. These multilayer societies are found in many species (reviewed in~\citep{Grueter2012MultilevelIssue}. The assumption behind these measures is that multi-level social systems are more complex than single-level societies, and that the more hierarchical layers, the more complex the social system.

\subsubsection*{Social roles}
The diversity of demographic or social roles in groups has been used as a measure of animal social complexity~\citep{blumstein1997does,Groenewoud2016PredationBreeders.,Anderson2001IndividualColonies}. Individuals can differ from each other via physical caste polymorphism, such as workers and drones in social insects, or via physiological and behavioral specialization, such as demographic roles where individuals can be divided into categories like breeding females or subadult males. The assumption of these measures is that increasing the number of different types or roles of individuals in a group results in increasingly complex social systems. 

\subsubsection*{Relationship differentiation} 
Relationship differentiation, or the number, heterogeneity, or diversity of relationship types can also be used to measure animal social complexity (\textit{e.g.}, ~\citep{Kutsukake2009ComplexityMammals,Fischer2017,bergman2015measuring, Weiss2019MeasuringModels}). These approaches assume that increasing the number of differentiated relationships or heterogeneity in social relationships  results in increasingly complicated social systems. Relationship types can be defined by researchers~\citep{Fischer2017,bergman2015measuring} or via mixture models from association patterns~\citep{Weiss2019MeasuringModels}. The assumption for these measures is that the higher the number of relationship types or the diversity of relationships, the greater the social complexity of the group.

\subsubsection*{Social uncertainty} 
Social uncertainty was recently proposed as a measure of social complexity~\citep{Ramos-Fernandez2018}. Their measure uses Shannon entropy to quantify the variability in subgroup composition in social structures with fission-fusion dynamics, where groups break apart and merge again, and where the particular individuals present in subgroupings may vary. This approach assumes that the predictability or uncertainty of an individual's social environment, and which individuals it is likely to be in contact with, can affect how dyadic relationships are regulated. The more uncertain an individual's social interactions are, the more cognitively difficult it may be for individuals to track and manage relationships, thus the greater the social complexity.

\section*{Three concepts from complex systems}

Several fundamental concepts from the field of complex systems can be useful for understanding social complexity. Here, we highlight three complex systems concepts that are particularly applicable to animal social systems: (1) \textit{scales of organization}, (2) \textit{compression}, and (3) \textit{emergence}. We highlight these concepts because they are individually useful in thinking about animal social complexity, and because they interact with each other in ways that provide a more unified perspective on social complexity. These concepts themselves are not measures of social complexity, nor in themselves necessarily indicative of high levels of social complexity, but rather should serve to guide thinking when using or developing approaches to complexity.\footnote{We would also like to point out that there is no unified definition of complexity in complex systems science. One long-standing debated point is whether randomness (or processes that appear random with respect to the system in question) have high complexity or not~\citep{grassberger2012randomness}. Recent developments lean towards ``not" and suggest that robust measures for quantifying a system's complexity should be based on the system's causal structure (\textit{e.g.,}~\citep{still2007structure}).} Incorporating these concepts into thinking about animal social complexity can add nuance to the standard measures of animal sociality. Below, we summarize each concept, summarize how it is used (or not) in existing approaches to social complexity, then discuss why it is important to incorporate into research on animal social complexity.

\subsection*{Scales of organization}

\begin{minipage}[t]{\textwidth}
\centering
\includegraphics[width=.85\textwidth]{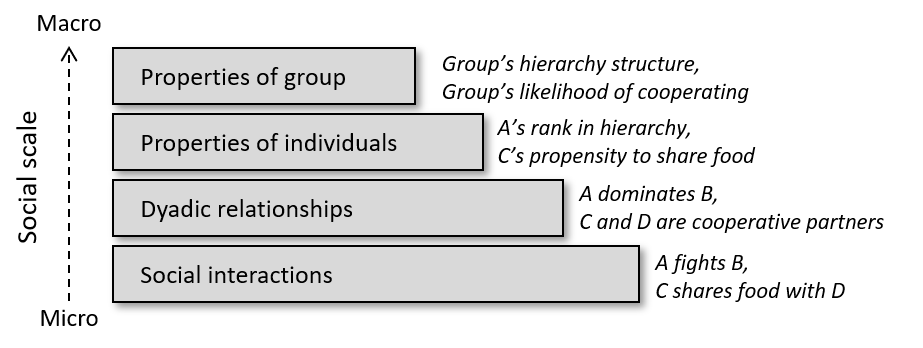}
\captionof{figure}{Illustration of commonly measured variables at different social scales, from micro to macro levels. Examples of summaries at each level are given in italicized text for two contexts: within group conflict and dominance hierarchies (upper examples) and food sharing and group cooperation (lower examples).}
\label{fig:socscales}
\end{minipage}  

Complex systems often have many scales, levels, and perspectives from which they can be described. This often yields different non-isomorphic decompositions of the system, meaning that the system can be described from several different structurally and functionally disjoint ways. A bottom-up perspective is used to understand the organization of many types of complex systems, in which the fundamental units of analysis are agents who take actions according to a set of rules~\citep{Macy2002FromModeling}. Animal social interactions can be summarized at different scales, from a micro level focused on how each individual experiences its social microcosm, to a macro level focused on a more global perspective of group-level social structures. 

In social systems, the agents are the individuals and the actions are interactions, events, and behaviors. Based on how individuals interact with each other at the micro social level, different types of macro-level social states can be produced. Social interactions and events become the foundation upon which more macro scale properties are built (\citep{Hinde1976InteractionsStructure}; Fig.~\ref{fig:socscales}). In this way, sociality is comprised of the micro-level of actions which aggregate to produce meso- and macro-level phenomena~\citep{Hinde1976InteractionsStructure,Little2012ExplanatoryBoat,Page2015}. Recent work in animal sociality has highlighted the importance of scale and has called for analyses of sociality --- especially measures of animal social complexity --- to clarify the scale on which analyses or measures are made, and to think carefully about which social scale is best for quantifying different types of complexity~\citep{Peckre2019ClarifyingComplexity,Aureli2019SocialGroups,Kappeler2019SocialEvolution}. 

To illustrate different types of sociality that occur at different social scales, we highlight two kinds of interactions: conflict and cooperation (Figure~\ref{fig:socscales}). At the most micro level, individuals can interact with each other by, for example, getting into fights or sharing food with each other. All of these interactions between specific individuals can be  summarized to describe the general relationship between pairs of individuals. For example, if we consider two individuals ``A'' and ``B'', if A wins many fights with B at the micro level, this can be summarized at the dyadic level as a dominant/subordination relationship. In a cooperative context, individuals ``C'' and ``D'' may consistently share food with one another on an event-by-event micro level, making C and D cooperative partners at the dyadic relationship level. For both conflict and cooperation, micro-level interactions that form the basis of pairwise relationships can be summarized across all the pairwise relationships in the group, thus forming a more macro-level social property for each individual. These summarized relationships can be categorical, for example if relationships are scored as affiliative or agonistic, or more continuous, for example if relationship strength is measured as the number of interactions among the pair of individuals. These individual social properties can represent rank in a dominance hierarchy (in the context of conflict) or an individual's propensity to share food (in the context of cooperation). In both cases, these individual-level social properties are built on the history of micro interactions and relationships. When these micro-scale social relationships change, it can cause changes in social properties at higher levels of organization, such as when shifts in aggression between individuals alters the structure of a larger group-level dominance hierarchy~\citep{Aureli2019SocialGroups}.

\subsubsection*{Benefits of incorporating scales of organization into social complexity}

There are several reasons why it is critical to consider scales of organization, particularly in descriptive approaches to animal social complexity.

First, social scales can be useful for making broad comparisons of sociality across species. Social systems can vary widely in their organization. This can make direct comparison of sociality difficult, as the biological relevance of different social features may be phylogenetically restricted within closely related species. One way around this issue is to use summaries or patterns, described at different levels of organization, to represent more abstract or high-level features of sociality. These abstraction can enable novel and far-reaching comparisons across different societies, which we discuss in more length in the next section on compression. It is especially important for descriptive approaches to complexity to be aware of, and explicit about, which scale is chosen to summarize sociality. For some research questions, it may be more relevant to summarize social interactions at a more micro level, while for others, summarizing at a macro level may  provide a better description of the social features of interest. This parallels with research on micro- and macroecology --- both are useful and interesting ways to summarize ecosystems on very different scales (\textit{e.g.},~\citep{Levin1992TheLecture,Maurer1999UntanglingPerspective,Brown1995Macroecology}). 

Second, attempting to compare across groups may be complicated if different groups have different relevant scales of organization. For example, in one group, the number of individuals may be a relevant measure, while in another, the number of family groups or matrilines is the important social feature, rather than the total number of individuals. Scales of organization can also be important because decision rules or other social mechanisms may operate on different scales. Recent evidence suggests that animals can use information at more than one social scale to make decisions, for example in which subgroup to join during foraging~\citep{Palacios-Romo2019UncoveringMonkeys}. As another example, experiments in baboons have demonstrated that they can classify others simultaneously by both that individual's rank in the dominance hierarchy (a property of individuals relative to a group) as well as by the individual's kinship relations (a more micro property at the dyadic level)~\citep{Bergman2003}. Looking at sociality on several different scales can allow researchers to detect the kinds and levels of information animals pay attention to, and to ask whether social features at different scales may be produced by different mechanisms. Analyzing systems on different scales is not limited to complex systems, but is particularly relevant to them because they tend to have organization at multiple scales. 

Finally, considering a complex system, such as an animal society, from multiple scales and perspectives can lead to a deeper and more multifaceted understanding of the system (e.g.,~\citep{wimsatt1972complexity,kauffman1976articulation}). In some cases, social features that would seem ``hard'' or complex if they had to be accomplished by individuals may be easier and less complex if these same features arise from a mechanism at a different scale (refer to  emergence section). Incorporation of social scale may then provide deeper insight into how information at different levels of hierarchical organization may feed into the social rules animals use. 

\subsubsection*{Existing use of scales of organization in social complexity measures}
There have been recent efforts to quantify social complexity using different scales of organization and to   think explicitly about sociality from the perspective of individuals within groups~\citep{Aureli2019SocialGroups}. Aureli and Schino proposed that group-level properties should be conceptualized as social complexity as ``seen from the outside'', versus more micro-level summaries at the individual level which quantifies social complexity as ``experienced from within''~\citep{Aureli2019SocialGroups}. This scale-dependent perspective is also referred to as bottom-up versus top-down~\citep{Weiss2019MeasuringModels}.   Bottom-up ``insider perspective'' approaches attempt to determine how each individual experiences sociality or the complexity of dyadic relationships, akin to a first person point of view. Top-down ``outside perspective'' approaches are similar to a ``bird's eye view'' of sociality, and treat social complexity as more of a network property at the group level~\citep{Aureli2019SocialGroups,Weiss2019MeasuringModels}. 

In most approaches to social complexity, measures focus on a single social scale to quantify complexity. For example, \textit{group size}, the number or heterogeneity of \textit{social roles}, and the extent of \textit{levels of structure} are all macro social properties measured at the level of the group. In contrast, \textit{relationship differentiation} is a more micro-level social property, as it is a summary of each individual's social relationships within the group. Relationship diversity or other individual-level social properties can be made more into more macro-level summaries if measures for each individual in the group are averaged across the entire group (\textit{e.g.},~\citep{Fischer2017}). Unlike many of the other measures, \textit{social uncertainty} implicitly includes more information about different social scales because it considers how individuals assort into different subgroups. These subgroups could be affected by both individual social properties and group-level social properties (Fig.~\ref{fig:socscales}), if they are somewhat stable and reform with some regularity. More recent work shows that fission-fusion decisions are based on two types of information: whether an individual is likely to be informed about food resources (partially determined by the individual's network centrality, a more macro-level social property of the individual) and the strength of the social relationship between individuals (a more micro-level social property of the dyadic relationship between the two individuals)~\citep{Palacios-Romo2019UncoveringMonkeys}. 

Complex systems contain multiple levels of organization, and therefore, multiple levels of description. This means that characterizing a complex system, or comparing multiple systems, is no easy feat. Within one animal social system, the amount of complexity we ascribe to it can differ depending on which level we have chosen to describe. When comparing multiple social systems, our evaluations of their relative complexity will depend on the levels at which each system has been described. Collecting information about the number of organizational levels and the complexity of each level of description puts researchers in a better position to compare social systems, because it allows them to access the space of possible comparisons that could be made rather than relying on one particular comparison.

%%%%%%%%%%%%%%%%%%%%%%%%%%%%%%%%%%%%%
\subsection*{Compression}
%%%%%%%%%%%%%%%%%%%%%%%%%%%%%%%%%%%%%

\textit{Compression} is an information-reduction process that  summarizes or abstracts patterns in observations~\citep{Marzen2017,Flack2017,Hankerson1998IntroductionCompression}. Compressing large amounts of observations can result in the removal of redundancies and noise, as well as relevant information (Fig.~\ref{fig:compression}). \textit{Coarse-graining} is a type of compression that results in a simplified representation of system behavior, generally through assigning continuously varying characteristics to bins or categories; once categorized in this way, the fine-grained details of items binned together are lost. 

\begin{minipage}[t]{\textwidth}
\centering
\includegraphics[width=1\textwidth]{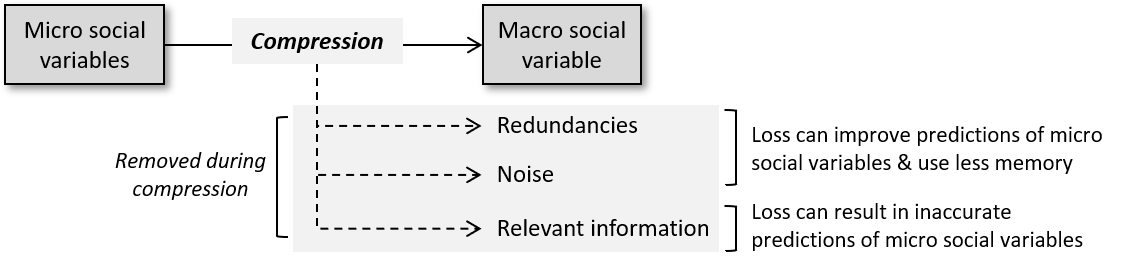}
\captionof{figure}{Compression of micro social variables into a macro social variable, indicating the types of information removed during compression and their potential consequences.}
\label{fig:compression}
\end{minipage} 

In a social setting, coarse-graining could involve grouping similar individuals into higher-order categories, and then utilizing information at the level of those categories rather than at the individual level. In the context of animal sociality, both the animals themselves and the researchers studying them compress social data. For animals, information is compressed when available information is filtered through the animal's perceptive or cognitive abilities or stored in a modified form in the animal's memory. For example, individuals may categorize others into clusters or sets based on matrilines while in other cases, the information on relatedness may be less compressed and more detailed if individuals need to differentiate relationships with others based on the degree of shared kinship~\citep{Aureli2019SocialGroups}). For researchers, compression occurs any time a body of observations are summarized with a measure or statistic.

\subsubsection*{Benefits of incorporating compression into social complexity}

Compression is important to consider when linking descriptive and causal approaches. Here, we summarize several reasons why compression is applicable to animal social complexity. 

First, compression and coarse-graining can result in information loss. As mentioned, when aspects of sociality are summarized, social observations, data, or information are compressed and some information in the original input is lost in the compressed output. Whether researchers use a descriptive or causal approach to social complexity, it is important to critically consider what kinds of social information may be lost during this process. This issue of information loss is relevant both to researchers seeking to summarize sociality and to the animals themselves, that must live within that society. In some cases, information loss during compression or coarse-graining can result in inaccurate predictions of micro and macro social variables if relevant information is removed (Fig.~\ref{fig:compression}). 

Second,  compression and coarse-graining can result in more accurate predictions of micro and macro social variables, for example when it removes noise and redundancies.  When dealing with finite sample sizes --- a situation faced both by researchers in the field and by animals acquiring social observations about their environments --- compression of information can help remove spurious correlations while retaining the ``real patterns''~\citep{Dennett1991,garlandPhD,joshua-pnp,garlandtdAIS}.  For this reason, as is known in statistics and machine learning, compression can improve generalization~\citep{bishop_pattern_2006}, which refers to the use of gathered information to make predictions about novel situations.  
For researchers using a descriptive approach to social complexity, this increase in predictability may allow for generalization to novel situations and other animal societies. For example, as we mentioned in the scales of organization section, macro-level coarse-grained social patterns may provide new opportunities for researchers to more easily and reasonably compare social properties across different groups. From a causal approach to social complexity, improved generalizations allows animals to use acquired information to make better predictions about future events, and to make better decisions in novel environments. 

Third, the type of compression or coarse-graining animals that use to process this information has important implications for perception, memory, and how social information is encoded. Individuals need to track certain kinds of information in order to make predictions and behavioral decisions, but in many cases, maintaining neural machinery required to perceive and process large amounts of information carries significant metabolic costs~\citep{levy_energy_1996,laughlin_metabolic_1998,laughlin_communication_2003,karbowski_thermodynamic_2009,kuzawa_metabolic_2014}. This forces individuals to make a trade-off between the cost of encoding and processing fine-grained data versus the cost of compressing that data and potentially making inaccurate decisions based on the coarse-grained representation~\citep{palmer2015predictive,Marzen2017}. There is a variety of evidence that demonstrates that animals detect and respond to coarse-grained information rather than perceiving all the fine-grain information potentially available. For example, zebra finches, which use color signals to select mates, recognize color categorically~\citep{Caves2018CategoricalSongbird}. This process reduces the fine-grained information about multiple color shades into two larger, more coarse categories, which may also encode coarse-grained information about the quality of potential mates. A similar coarse-graining process occurs when animals respond categorically to vocalizations from individuals from other dialects (\textit{e.g.},~\citep{Baker1985TheDialects,Wright2018VocalEvolution,Henry2015DialectsFunctions}), or when animals change their behavioral responses when dealing with neighbors versus strangers (\textit{e.g.},~\citep{Ydenberg1988NeighboursAttrition,Temeles1994TheEnemies}). When using a causal approach to animal social complexity, it is important to critically evaluate different ways of compressing or coarse-graining social observations. For this approach, a ``good'' compression or coarse-graining under is often one that extracts patterns that are relevant for animals' survival, fitness, or short-term goals. Such patterns may or may not be perceivable by the animals themselves. However, it may be evolutionarily adaptive for animals to be ``tuned'' to those coarse-grained variables which are particularly relevant to their biological goals. Such tuning may be carried out either via particular sensorimotor capabilities~\citep{palmer2015predictive}, or via higher-level neural systems which compress incoming information. Additional empirical work on perceptual and cognitive abilities of animals would be required to properly understand the comppresion performed by a particular species.

Finally, compression and coarse-graining are critical to consider when estimating the cognitive load of sociality. Compressed information requires less memory and processing bandwidth, and, as mentioned, animals may have evolved to either perceive information in a compressed way or may have evolved cognitive abilities which allow them to compress information. 
This potential reduction in the cognitive load of sociality is important to consider, especially in its implications for the  idea that
sociality is increasingly difficult to cognitively manage as the number of individuals, the number of relationships, and the types of relationships increase. Previous research has suggested that sociality would be ``combinatorially explosive'' if individuals need to track all the dyadic or triadic relationships in larger groups~\citep{Seyfarth2015,Seyfarth2001CognitiveMonkeys}. However, if individuals can appropriately compress information about their social worlds, then the relationship between the total amount of available information and the cognitive load of sociality may be quite nonlinear. In particular, a society that appears at first glance to be complex or cognitively challenging to track may actually be much more simple. An example of this kind of ``missing social knowledge'' has been documented in gelada, where individuals live in large spatially-organized groups, but in fact only recognize individual conspecifics that belong to their immediate social group~\citep{bergman2015measuring}. In this case, gelada appear to compress the available social data into coarse-grained categories: within-family individuals in one category and all others lumped into another category, resulting in lower social complexity than was originally thought.

\subsubsection*{Existing use of compression in social complexity measures}

Existing measures of animal social complexity implicitly incorporate some aspects of compression and coarse-graining, but most do not explicitly address why compression is an important concept to consider. For example, any single-value measure or summary statistic of animal sociality is a compressed description of the social organization, but these summaries can vary in the degree to which social information is compressed and which types of information are retained in the summary.  

Summary statistics, or single-value measures of social complexity, mainly acquire explanatory or causal relevance --- and thus becomes a better animal social complexity measure --- when they align with how the individuals or groups themselves coarse-grain their worlds. For example, for an organism (like humans) that can independently recognize and maintain relationships with other conspecifics, \textit{group size} is probably an informative measure of social complexity, but for an organism for whom other conspecifics are an undifferentiated mass (like some schooling fish), or divided into general categories (like caste in social insects), the raw number of conspecifics does not  reflect the same thing. Measures like \textit{relationship differentiation} and \textit{social roles} retain categorical coarse-grained information about social roles and relationships in groups, but information about the number of individuals in the group is compressed out. 

However, it can be complicated to bin roles or relationships into biologically meaningful categories that correspond to how animals themselves perceive their social worlds. For example, researchers may categorize roles or relationships much more finely than the animals themselves, which may lump several of these apparently different roles or relationships into the same larger category. The reverse situation is also possible, where animals perceive much finer-grained roles or relationship types do the researchers in their analysis. In either case, a mismatch in the granularity at which researchers and animals coarse-grain the social roles or relationships can strongly affect  measures of social complexity. This same issue may affect measures of \textit{social uncertainty}, which should take into account the perceptive and recognition abilities of the species of interest. Failing to take perception, recognition, and memory into account when quantifying social uncertainty may result in over-estimating the uncertainty any one individual may experience, especially if for example, all non-relatives are coarse-grained into a single amorphous group and recognized as a category rather than different individuals. 

In all these cases, these social complexity measures are used with the hope that they can provide us with a useful summary of different types of sociality, but most do not address the causal mechanisms underlying social patterns. Existing methods apply a ``more is harder'' approach: the more individuals to recognize and remember, the more relationship types to maintain, the more roles to fill, the more levels of society, the more uncertain the social groupings, the more complex the society. However, what we as researchers perceive as ``more'' may not align with how animals perceive their social worlds, especially if the animals compressing or coarse-graining information in ways that are not obvious to researchers studying those systems.

To summarize, compression is an operation that can be performed by both researchers attempting to understand a complex social system and animals attempting to navigate their own complex social system. The compression choices that researchers make can be positive, leading to better predictions and generalizations, or negative, leading to the loss of relevant information about a particular level in the system. The compression and coarse-graining that animals do are causally relevant to the functioning of their own complex social system. Researchers can obtain more accurate complexity measures when their coarse-graining choices align with those used by the organisms in the social system.

%%%%%%%%%%%%%%%%%%%%%%%%%%%%%%%%%%%%%
\subsection*{Emergence}
%%%%%%%%%%%%%%%%%%%%%%%%%%%%%%%%%%%%%

Systems with a many interacting components, such as animal societies, can exhibit \textit{self-organization}, which occurs when local interactions between components give rise  macro-level order~\citep{Smaldino2014TheTraits, Miller2009ComplexLife,Hinde1976InteractionsStructure}. Classic examples of self-organization include bird flocking and fish schooling, in which large groups of animals move in a coordinated manner, even though each animal is believed to only directly interact with its immediate neighbors~\citep{hemelrijk2012schools}.  Self-organization is often associated with \textit{emergence}, in which the large-scale order exhibits qualitatively new types of properties. To use the previous example, large flocks of birds exhibit emergent  macro-level properties such as propagating ``waves'' of density or orientation, which transfer information between distant parts of the flock~\citep{procaccini2011propagating}.  Wave-like dynamics are an example of an emergent property, since they doesn't exist (and cannot even be meaningfully defined) at the level of individual birds. This aspect of novelty differentiates emergent properties (such as ``wave-like behavior'') from simple aggregative properties (such as the fact that collectives have larger mass or number than individuals~\citep{wimsatt1986forms}).

Emergent phenomena can arise from simple behavioral heuristics at the micro level, like those documented in animal movement and problem-solving studies (\textit{e.g.},~\citep{Berdahl2013EmergentGroups.,Strandburg-Peshkin2015SharedBaboons}). Emergent patterns can also be information-based, such as rank information contained within dominance hierarchies in many species~\citep{hobsondedeo2015,Hobson2018StrategicKnowledge,BradburyVehrencamp2014,Aureli2019SocialGroups}, culturally-learned behaviors (\textit{e.g.},~\citep{Aplin2018CultureEvidence,Whitehead2017Gene-cultureDolphins.}), or leadership in collective movements (\textit{e.g.}, ~\citep{garland-leadership,Torney2018InferringCaribou.}).

An important aspect of many complex systems is that macro-level properties which emerge from micro-level interactions can then ``feedback'' and constrain   micro-level interactions and dynamics~\citep{Miller2009ComplexLife,Page2015}, a process sometimes called \emph{downward causation}~\citep{Flack2017,Campbell1974DownwardSystems}. To use the previous example, when a flock of birds is collectively moving in a given direction (the macro-level emergent property), the orientation of individual birds becomes constrained by the macro-level collective movement, and it becomes difficult for them to deviate their individual flight paths.  
Downward causation is especially pervasive in systems like animal societies, which are composed of cognitively sophisticated individuals that can sometimes explicitly recognize and use macro-level properties to direct their individual behavior.  It is critical to understanding the link between aggregated social knowledge at macro scales and its effect on micro level interaction decisions, like how some animals use emergent rank information to choose who to fight at the micro level~\citep{hobsondedeo2015,Hobson2018StrategicKnowledge}. 
The feedback effect of social context on behavior is sometimes conceptualized in animal behavior as an \textit{audience effect} (\textit{e.g.},~\citep{Coppinger2017StudyingResearch}), but can be much more broad and unrelated to communication. For example, feedback from social context is critical to fanning behavior in honeybees, which are less likely to fan when solitary, as compared to when they are in a larger group~\citep{Cook2013SocialHoneybees}.

We illustrate the combined processes of emergence and downward causation  in social systems in Fig.~\ref{fig:feedback}, using examples from the interactions between aggression and dominance hierarchies and affiliation and cooperation. The social system self-organizes across different social scales by incorporating micro-level interactions into macro-level properties (upwards-directed arrows). These more macro-level social properties can then affect future micro-level interactions through downward causation, indicated by the outwards and downwards facing arrows. As these macro properties are forming in the system, downward causation serves to give structure to the micro-level interactions.

\begin{minipage}[t]{\textwidth}
\centering
\includegraphics[width=1\textwidth]{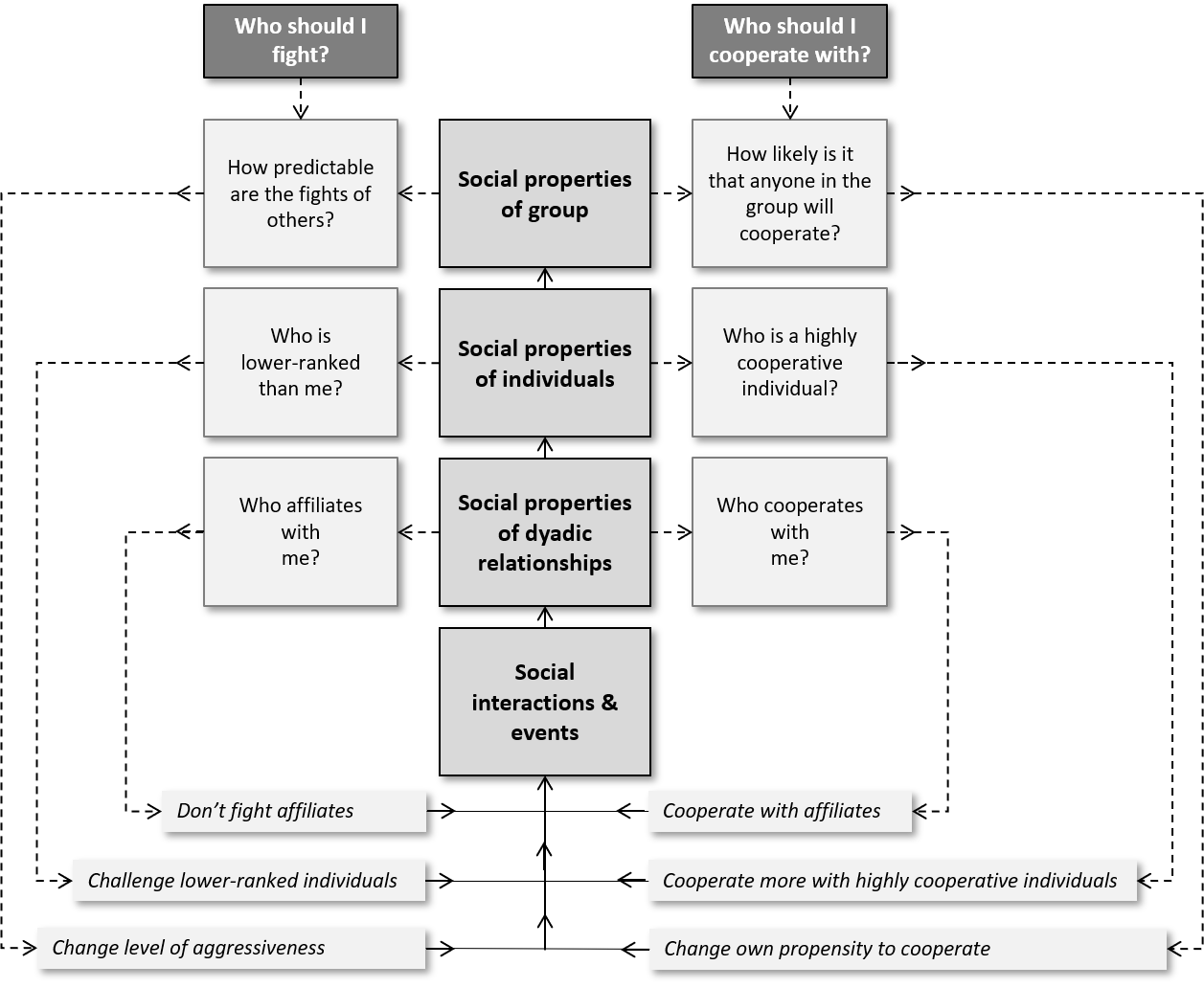}
\captionof{figure}{Examples of emergence and downward-causation across social scales in two social contexts: fighting decisions (left) and cooperating decisions (right). Upward-pointing arrows indicate how micro-level information becomes aggregated into more macro-level social properties (in some cases, these represent coarse-grained information); downward-pointing arrows indicate feedback via downward-causation. The gray boxes with white text indicate two different behavioral decisions individuals may have to make (who to fight or who to cooperate with). The light gray boxes indicate the type of information individuals may use to make those decisions depending on which social scale the information is drawn from. The central gray boxes depict the social scale and the italicized text indicates how the individual incorporates information into its behavioral decisions. The dashed lines indicate micro-level decision-making based on macro-level information, while the solid lines indicate how the effects of those decisions percolate up from micro to macro levels and are aggregated at the more macro levels.}
\label{fig:feedback}
\end{minipage}

\subsubsection*{Benefits of incorporating emergence into social complexity}

The notions of emergence and downward causation  can help us understand how the decisions that animals make at a micro-level may be contingent on the rules, information, or knowledge at a more macro-level~\citep{Flack05072012,hobsondedeo2015}. This is especially relevant for  researchers interested in the causal aspect of  animal  social complexity for several reasons.

First, incorporating ideas about emergence could help better understand micro to macro connections, and how micro-level interactions may lead to macro-level patterns.  Emergence can be used as a causal explanation for why multiple scales of organization are present in a system such as an animal society. For example, aggression in monk parakeets appears to be largely based on their own emergent information about rank in the dominance hierarchy~\citep{hobsondedeo2015}. Emergent properties can also play the important role of  stabilizing a system's dynamics: macro-level properties often change on a slower time scale than micro interactions, thus downward causation from slowly changing macroscopic social properties can serve as a stabilizing force, helping  societies balance the trade-off between maintaining robust information about social interactions while still responding adaptively to novel perturbations and events~\citep{Flack05072012,Flack2017}.

Second, the idea of emergence and downward causation has important implication for the ``more is harder'' motivation for studying social complexity. These notions suggest  that some behaviors seem complex but can arise from rather simple small-scale rules, thus they are not as complex as we might naively think. For example, ant colonies excel at collectively finding optimal routes to nearby food resources or new nests (e.g.~\citep{beekman2001phase,sumpter2005principles}. They do this by simple local rules of interaction between individual ants, and such rules may in fact be quite simple to evolve.  In the context of bird sociality, as mentioned above, it is thought that  dominance hierarchies arise easily if the birds have some way to measure their own rank in a society and then take actions based on that information, a type of downward causation mechanism\cite{hobsondedeo2015}. 
The opposite can also potentially occur, where some apparently simple global properties may actually be much more difficult to achieve with a group of distributed individuals than might appear at first glance. Emergence and self-organization has important implications for evolutionary analyses of animal social complexity. Social phenomena which may appear to require the evolution of specialized mechanisms can, in some sense, arise ``for free'' as a product of  self-organization~\cite{kauffman1993origins}; social phenomena which may appear to require no evolutionary explanation at all may, in fact, do so when carefully analyzed.

Finally, from a descriptive point of view, emergence implies the presence of multiple scales of organization, where each scale has its own properties and rules. Thus, carefully thinking about the appropriate scale of analysis (as discussed in the scales of organization section) is particularly important for systems that exhibit emergent properties.

\subsubsection*{Existing use of emergence in social complexity}
While self-organization and emergence are often invoked to describe social patterns like flocking birds or schooling fish, the idea of emergent properties has not been incorporated into relationship-based approaches to social complexity. Measures like group size, relationship differentiation, levels of structure and  uncertainty in social dynamics exclusively focus on social patterns at the micro levels and do not address macro level emergent properties of the social system. None of the five social complexity measures we summarized earlier explicitly incorporate ideas about the potential for downward causation, even though this could  be a critical component in cognitively-based explanations of social complexity. 

Measuring complexity as social uncertainty does touch self-organization of groups, although  it does not explicitly incorporate downward causation, which may be an important process to consider. For example, animals may respond to social uncertainty by changing their fission-fusion patterns over time and altering how they move between subgroups, so as to move preferentially with a certain subset of individuals. This kind of biased movement may be one way that animals could respond to via a feedback mechanism and reduce social uncertainty, leading to greater self-organization, and causing a reduction of social uncertainty over time. Thinking about how complexity may vary temporally is important because social dynamics can cause their apparent complexity to go up or down over time, which could further complicate comparability between social systems. 

To summarize, emergence is a hallmark of complex systems such as animal societies, and can also lead to downward causation. Because simple micro-level rules can give rise to macro-level emergent behaviors in a social system,  organisms in a complex social system need not necessarily display complex individual behavior, showing yet another reason why ``more'' is not necessarily ``harder''.    Self-organization via emergence is also an alternative to an evolutionary explanation to understand how and why complexity arises and is maintained in animal social systems.

%%%%%%%%%%%%%%%%%%%%%%%%%%%%%%%%%%%%%
\section*{Conclusions}
%%%%%%%%%%%%%%%%%%%%%%%%%%%%%%%%%%%%%

We summarized several ways in which animal social complexity is measured, and described how these measures incorporate (or fail to incorporate) three foundational concepts from the science of complex systems approaches: scales of organization, compression, and emergence. All three of these key concepts can be applied together to better understand various aspects of social complexity, and what aspects of complexity are included or excluded from any particular measure. If accounting for one of these concepts seems critical to the system of interest, but is not considered by the method used to evaluate social complexity, the researcher should consider an alternative method. 

Future work on animal social complexity can benefit in many ways from incorporating these foundational concepts. In particular, we think that these concepts can help researchers better think about sociality from multiple perspectives, and to clarify the motivation for quantifying complexity in the first place: is it so that researchers can more easily compare systems to one another using a descriptive approach, or is it to understand how animals themselves create and understand their social worlds, using a causal approach to understanding mechanistic processes?

A major driver of many comparative social studies is to gain insight into the evolution of sociality. To do this, it is critical to be able to compare sociality across different species, to trace evolutionary trajectories, determine selection pressures, find interactions with other traits like cognitive abilities, and identify cases of convergent evolution. However, it is not an easy task to identify an approach to sociality that is biologically relevant, and which allows comparisons of species with potentially very different social systems, making quantifying social complexity a difficult task. For the purposes of making  phylogenetically-broad comparisons of the evolution of sociality, a successful measure is one that can be applied to a wide range of species in a biologically meaningful way, can detect similarities across these various species, and that can be quantified using available datasets or new data that is feasible to collect. 

Animal social complexity is, at its heart, a multi-faceted topic. It is short-sighted to expect that a single approach to complexity will capture all the relevant aspects of social complexity in all research contexts.   Conceptualizing complexity along multiple axes helps circumvent some of the drawbacks of quantifying social complexity with a single measure, and a multidimensional approach to social complexity, for example as seen in some recent proposals~\citep{Lukas2018, Rubenstein2017SocialEvolution}, is likely to improve comparisons of social systems when taking a ``different is interesting'' descriptive approach. However, a persistent issue with  multidimensional approaches is that if the separate axes of complexity do not individually reflect relevant measures of complexity, it is unlikely that looking at the combination of these measures will provide additional insight into social complexity, or a better description of the sociality. Thus, appropriate axes must be chosen carefully. 

A major short-coming of many current animal social complexity measures is that they are fundamentally descriptive, and thus cannot directly address causal questions. We think that investing more research effort into uncovering the rules underlying social interactions is an important way forward. New methods have been recently proposed that look more at how animals use information at more macro scales to structure rules that inform their micro-level interactions~\citep{Hobson2018StrategicKnowledge}.

In this paper, we provide readers with a suite of conceptual tools for critically thinking about different approaches to animal social complexity, as well as important points to keep in mind  when evaluating or designing complexity measures. Ultimately, researchers studying animal social complexity should carefully balance the feasibility of obtaining data with the biological relevance of their analysis, keeping in mind the ultimate goal of understanding how animal sociality is structured, what causal factors lead to differences in sociality, and how different types of sociality evolved or emerged.

\section*{Acknowledgements}
The authors thank the Santa Fe Institute for helping to support this research and thank Rafael Lucas Rodriguez for organizing the ``Unasked Questions'' symposium at the 2018 Animal Behavior Society meeting as well as inviting us to be part of this special issue. EAH was supported by a Complexity Fellowship from the ASU-SFI Center for Biosocial Complex Systems. VF and JG were supported by Omidyar Fellowships at the Santa Fe Institute.  VF was also supported by the University of Melbourne's Research Fellowship in Computational Cognitive Science.

\newpage

%\bibliography{FC}
\bibliographystyle{plainnat}
\bibliography{MASTERBIB2}
%\bibliography{MASTERBIB}
\end{document}